# Achieving interoperability between the CARARE schema for monuments and sites and the Europeana Data Model

Valentine Charles
Antoine Isaac
Europeana Foundation,
the Netherlands
valentine.charles@kb.nl
aisaac@few.vu.nl

Kate Fernie
MDR Partners, UK
kate.fernie@mdrpartners.com

Costis Dallas
Dimitris Gavrilis
Stavros Angelis
Digital Curation Unit-IMIS
Athena Research Centre, Greece
c.dallas@dcu.gr
gavrilis@gmail.com
s.angelis@dcu.gr

**Abstract**

Mapping between different data models in a data aggregation context always presents significant interoperability challenges. In this paper, we describe the challenges faced and solutions developed when mapping the CARARE schema designed for archaeological and architectural monuments and sites to the Europeana Data Model (EDM), a model based on Linked Data principles. The purpose of this mapping was to integrate more than two million metadata records from national monument collections and databases across Europe into the Europeana digital library.

## 1 Introduction

Since its inception in 2008, the Europeana.eu digital library has grown to become a major information source on European material and intangible cultural heritage, integrating and providing access to a wide spectrum of cultural objects ranging from books, manuscripts and archival items to visual artworks, archaeological artefacts and records of local history and culture. Europeana provides access to authenticated, reliable metadata records from more than two thousand cultural heritage institutions across Europe, made available through the mediation of a large number of national and thematic aggregators.

Large-scale information integration of heterogeneous information in the cultural heritage domain, such as performed by Europeana, introduces significant interoperability challenges. Crosswalks and mapping between different data models and schemas is widely recognized as a necessary step in achieving information integration (Papatheodorou, 2012). One strategy for dealing with such challenges has been by way of thematic aggregation of digital resources, ensuring a degree of homogeneity before metadata are made available for harvesting and ingestion into Europeana. *CARARE – Connecting Archaeology and Architecture for Europeana*, a three-year EU-funded project that ran from February 2010 to January 2013, was established with the intent of aggregating content from the archaeology and architecture heritage domain and providing it to Europeana. CARARE aimed to ensure a high degree of homogeneity and quality across widely diverse collections, including those of heritage agencies, ministries, museums and archives (Hansen & Fernie, 2010).

Work on the CARARE project coincided with a period of significant change, in which Europeana initiated a process of evolution from using the relatively minimal *Europeana Semantic Elements* (ESE) metadata standard to the significantly more expressive, and information-rich, *Europeana Data Model* (EDM, 2012). During 2012 and early 2013, the twenty nine partners of the CARARE project delivered over two million records about archaeological sites and historic monuments from all across Europe to Europeana in the new EDM format. The architecture adopted by CARARE was based on aggregating and delivering metadata through MoRe (Monument Repository), an OAIS-compliant, preservation-grade, curation oriented thematic aggregator managing the full workflow from ingestion to Europeana harvesting and ingestion (Papatheodorou et al., 2011a; Gavrilis et al., 2013).

This paper presents strategies and solutions developed to resolve challenges related to the integration of CARARE monuments and sites metadata into Europeana. In the following sections, it introduces the CARARE schema aimed at ensuring maximal retention of information on archaeological and architectural monuments and sites, and their representations; it outlines the rationale and main properties of EDM; it presents the key challenges faced in mapping CARARE into EDM metadata; and, finally, it illustrates how this approach contributed to a successful process of integrating more than two million metadata records on immoveable cultural assets, and their digital representations, and making them accessible through the Europeana digital library.

## 2 The CARARE schema

CARARE has established a domain-specific metadata schema (Papatheodorou et al., 2011b) aiming at the representation of archaeological sites, historic buildings and monuments. This metadata schema makes use of established standards from the archaeology and architecture domain. In particular the CARARE schema adopted the main structure of the MIDAS Heritage standard (MIDAS, 2012), enriched with elements from the POLIS DTD (Constantopoulos et al., 2005), and the LIDO schema (Coburn et al., 2010), in order to capture information about individual objects, buildings and sites but also relevant spatial and temporal information. The underlying conceptual foundation used to model relationships between different entities defined by the CARARE schema was the CIDOC CRM standard (Crofts et al., 2003). The CARARE schema has been designed to allow content providers to map their source data to a rich schema while minimizing the loss of semantics.

Collections contributing to the MoRe metadata aggregator ranged from archaeological and architectural monuments and sites databases to digital libraries of heritage-related visual resources. The mission of CARARE was, therefore, to integrate information about immoveable heritage objects, other cultural objects depicting such heritage objects (such as paintings, drawings, plans and 3D models), and related digital resources, while representing adequately their relationships, as illustrated in Fig. 1.

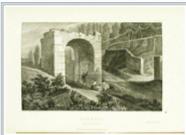

FIG. 1. A heritage asset and other objects related to it on the CARARE landing page:
http://store.carare.eu/landing-page-ha.php?id=iid:2920150&eid=HA:6161

The CARARE schema is based on four major concepts that are wrapped into a main entity—the CARARE record:

- One *Heritage Asset* which holds the metadata for a site or a monument. This entity includes descriptive and administrative metadata. This entity is unique in the CARARE record.
- At least one *Digital Resource* which holds the metadata about a digital resource; such as digital representations of books, photographs or 3D models. In an aggregation context, this entity describes the digital objects which are made accessible to Europeana.
- Optional *Collection level descriptions*.
- Optional *Activities* holding metadata about an event or an activity, such as an excavation.

In addition the CARARE schema allows, by means of specific element values, identification of additional place and agent entities, which can be used to document contextual information related to a specific site or monument. CARARE entities can be explicitly related through appropriate element values to denote a relationship between heritage assets (for instance, in the case one asset is part of another asset, which is typical in architectural and archaeological heritage), or a relationship between a heritage asset and a digital resource that represents the asset.

## 3  The Europeana Data Model

Europeana providers create their original metadata in dozens of heterogeneous formats and vocabularies. Until recently, all data had to be mapped to a simple common-denominator vocabulary, the Europeana Semantic Elements (ESE). ESE mostly uses fields from Dublin Core (DC) in combination with simple string values, in flat records. This raises many quality issues, related to the merging of data on original object and digitized version and aggregation process into one record, the impossibility of expressing richer data in a machine-readable form, or of expressing semantic links between objects with other objects or persons, places, concepts, etc.

On these grounds, Europeana devised a new Europeana Data Model, EDM, for its data aggregation and dissemination processes. EDM draws from the vision of the Semantic Web and Linked Data and from rich metadata models like CIDOC CRM. It notably enables the use of Dublin Core properties with fully-fledged resources instead of only as strings, and includes new properties (e.g., edm:isDerivativeOf, edm:isRepresentationOf) to link objects together (e.g., for composite objects with parts) or connect them to external resources. The aim is to allow providers to provide Europeana with richer data.

EDM provides and re-uses many classes and properties from the following namespaces: Resource Description Framework (RDF), OAI Object Reuse and Exchange (ORE), Simple Knowledge Organization System (SKOS) and Dublin Core (DC). In this section, we focus on the main types of resources that Europeana providers should most consider while mapping their data to EDM:

- *Provided Cultural Heritage Objects* (CHOs, edm:ProvidedCHO) denote the original objects—either physical (painting, book, etc.) or born-digital (3D model), which are the focus of description and search in Europeana. The choice in granularity of description chosen for the ProvidedCHO belongs to the data provider, within the limits of relevance set by Europeana.
- *Web Resources* (edm:WebResource) represent a digital representation of the provided cultural heritage object, published on the web.
- *Aggregations* (ore:Aggregation) group the ProvidedCHO and the WebResource(s) into one bundle, where information on the aggregation process is also recorded (e.g., the provider of the data).
- EDM defines contextual resources that can be used to provide more information related to the object (e.g., edm:Agent, edm:Place, edm:Concept, edm:TimeSpan).

## 4   Key mapping decisions for converting from CARARE to EDM

The MoRe metadata aggregator maintains an OAIS compliant metadata repository and supports transformation of CARARE records into EDM metadata suitable for harvesting and ingestion by Europeana, implemented by means of XSLT stylesheets (Papatheodorou et al., 2011a; Gavrilis et al., 2013). The challenges faced in achieving a transformation of CARARE schema into EDM metadata were of conceptual, rather than technical nature. An initial mapping from the CARARE schema to EDM was developed as early as 2010 and was significantly updated in a collaborative process involving the CARARE and Europeana teams, based on iterative trials of converting sample batches of CARARE metadata into EDM through 2012. This section presents the key considerations and decisions made in the course of this mapping process.

### 4.1   Aligning CARARE entities with EDM classes and establishing object identity

The mapping uses EDM classes to describe information about:

- Immoveable heritage assets, such as monuments, buildings or other real world objects, identified by a set of particular characteristics that refer to their identity, location, related events, etc. Information carried by a *Heritage Asset* includes textual metadata (such as title, etc.), thumbnails and other digital objects.
- Other real-world cultural objects with digital representations, which provide sources of information about the *Heritage Asset*: historic drawings and photographs, publications, archive materials etc.
- Born-digital resources related to these objects, such as 3D models.

The mapping to EDM produces a edm:ProvidedCHO and an ore:Aggregation for each *Heritage Asset*, the ProvidedCHO carrying the data that pertains without ambiguity to the original object, while the Aggregation provide information that results from the process of producing and publishing data (including digital content).

Each CARARE *Digital Resource* leads to the creation in EDM of one edm:WebResource that will be bundled with the edm:ProvidedCHO generated from the *Heritage Asset*, in its ore:Aggregation. Potentially each *Digital Resource* may also produce one separate edm:ProvidedCHO and ore:Aggregation according to the scenarios detailed in Section 3.3.

The Europeana Data Model requires a unique identifier for each resource. Since unique identifiers could not be guaranteed in the original data submitted by content providers, unique identifiers were created within the CARARE system for each ore:Aggregation resource, based on the local identifier provided by content providers (Fig 2).

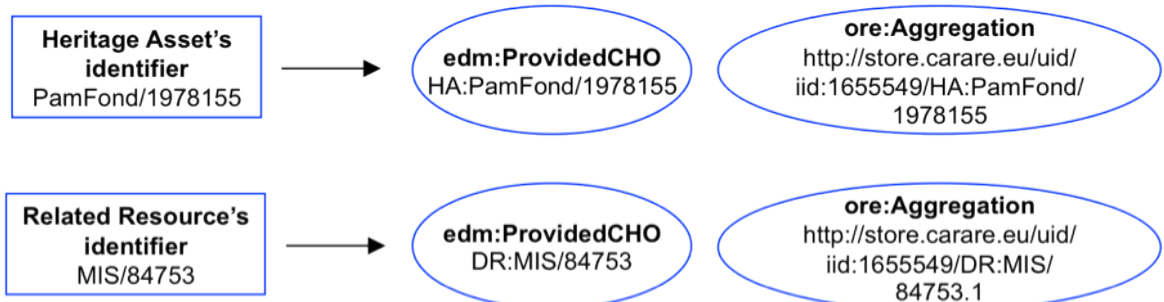

FIG. 2.  Creation of EDM resource identifiers from local identifiers.

These aggregation identifiers are "Web-enabled", in the sense that they redirect to a landing page that CARARE creates for each object. The landing page not only provides a stable unique identifier but also visually aggregates the *Heritage Asset* with its representations in *Digital Resources*, *Collection* information and information about related *Activities*, thus providing essential context for the interpretation of underlying relations to users.

Other EDM resources created from CARARE data are provided with unique identifiers derived from local identifiers in a way similar to the one described above. WebResources have necessary HTTP identifiers as they belong to the Web. ProvidedCHO identifiers are not all web-enabled, as this is not a crucial requirement for them. The following example illustrates our approach:

```
<ore:Aggregation
rdf:about="http://store.carare.eu/uid/iid:1655013/DR:MIS/161379.3">
  <edm:aggregatedCHO rdf:resource="DR:MIS/161379"/>
  <edm:dataProvider>Národní památkový ústav / National Heritage Institute
  </edm:dataProvider>
  <edm:provider>CARARE</edm:provider>
  <edm:isShownBy
rdf:resource="http://iispp.npu.cz/mis_public/documentPreview.htm?id=161379"/>
  <edm:rights ref:resource="http://creativecommons.org/licenses/by-sa/3.0/"/>
</ore:Aggregation>
```

### 4.2   Representing archaeological objects and cultural objects related to them

The different information sources related to an archaeological asset (as in Fig. 4) give rise to different ProvidedCHOs. Each print, map, or book about an archaeological place, such as a book on a house in Pompeii, counts as a separate object provided to Europeana. The new identifiers assigned to each object make it possible to create an explicit link between the heritage asset and a related object. For instance, when a document (as ProvidedCHO in EDM) represents an heritage asset (e.g., a print showing a monument) the relation is expressed using the property edm:isRepresentationOf with the URIs of the corresponding resources.

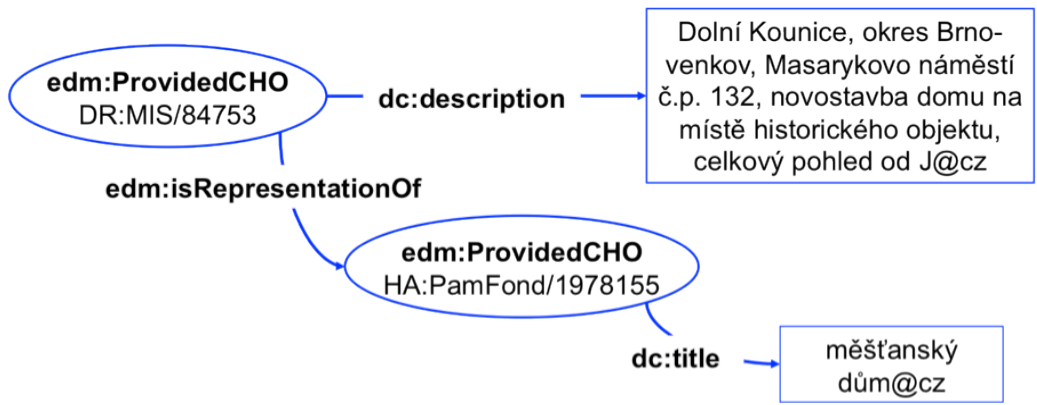

FIG. 3.  Relation between an heritage asset and a cultural object indicated by edm:isRepresentationOf.

### 4.3   Deciding what counts as a Europeana object

As mentioned above, the determination of an edm:ProvidedCHO submitted to Europeana belongs to the data provider. Clearly, it is appropriate to represent CARARE heritage assets as Provided CHOs. Making the decision for other kinds of cultural resources is more difficult, as the grain of *Digital Resources* vary greatly across the CARARE datasets. Decisions on how to deal with this issue were made on the basis of the specific situation governing individual datasets, falling under the following scenarios:

- Scenario 1: Each CARARE record contains a heritage asset such as a building. Often a single *Heritage Asset* building may be related to a number of derivatives of cultural objects, which, from an archaeological point of view, offer different views on this heritage asset, such as historic drawings, books and photographs. When converting such a CARARE record to EDM, the *Heritage Asset* and each of the cultural objects representing it give rise to individual ProvidedCHOs. The edm:isRepresentationOf property is used to link the ProvidedCHO resulting from a *Digital Resource* to the ProvidedCHO of the *Heritage Asset* it represents, as shown in Fig.4.

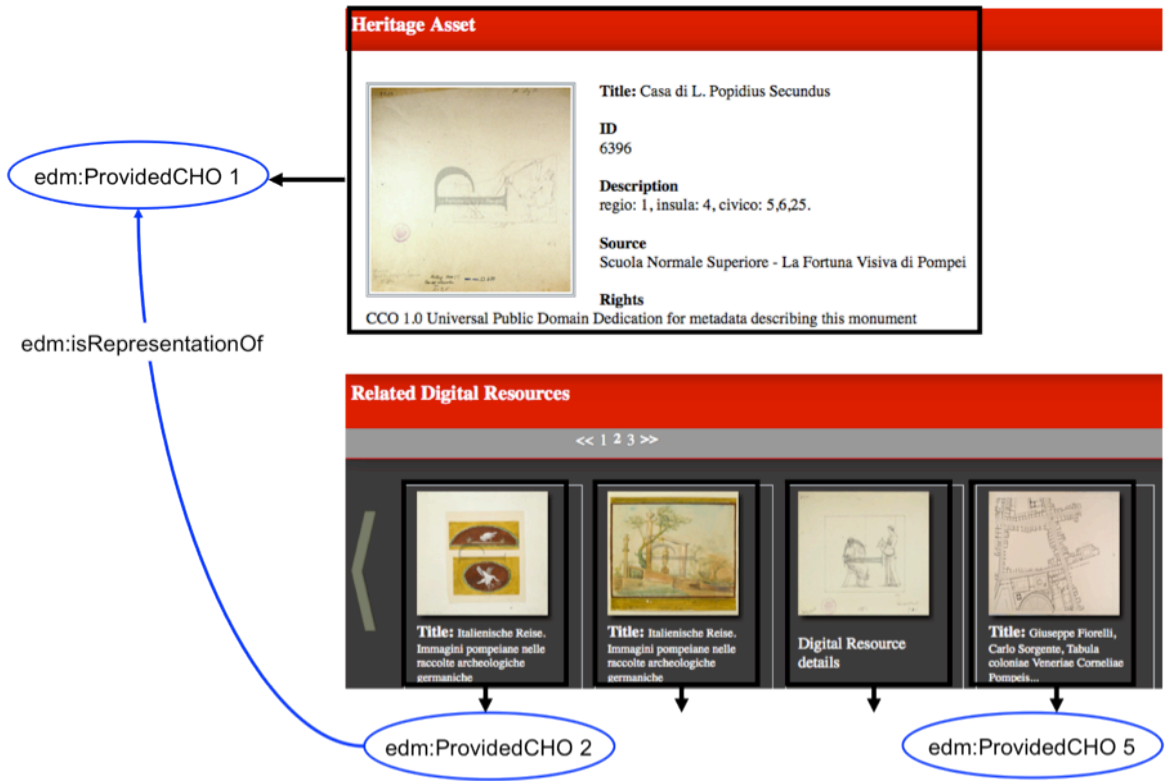

FIG. 4. Scenario 1: CARARE *Heritage Asset* and *Digital Resources* converted into EDM Provided CHOs.
URI ProvidedCHO 1: http://store.carare.eu/landing-page-ha.php?id=iid:2920154&eid=HA:6396

- Scenario 2: a book, photograph or map relates to more than one *Heritage Asset* and is contained in more than one CARARE object. As defined in scenario 1, each *Heritage Asset* creates a ProvidedCHO and each derivative cultural object representing it also creates a ProvidedCHO. To ensure non-redundancy, any duplicate Provided CHO created as a result of the book or picture being referenced by more than one *Heritage Asset* is removed.

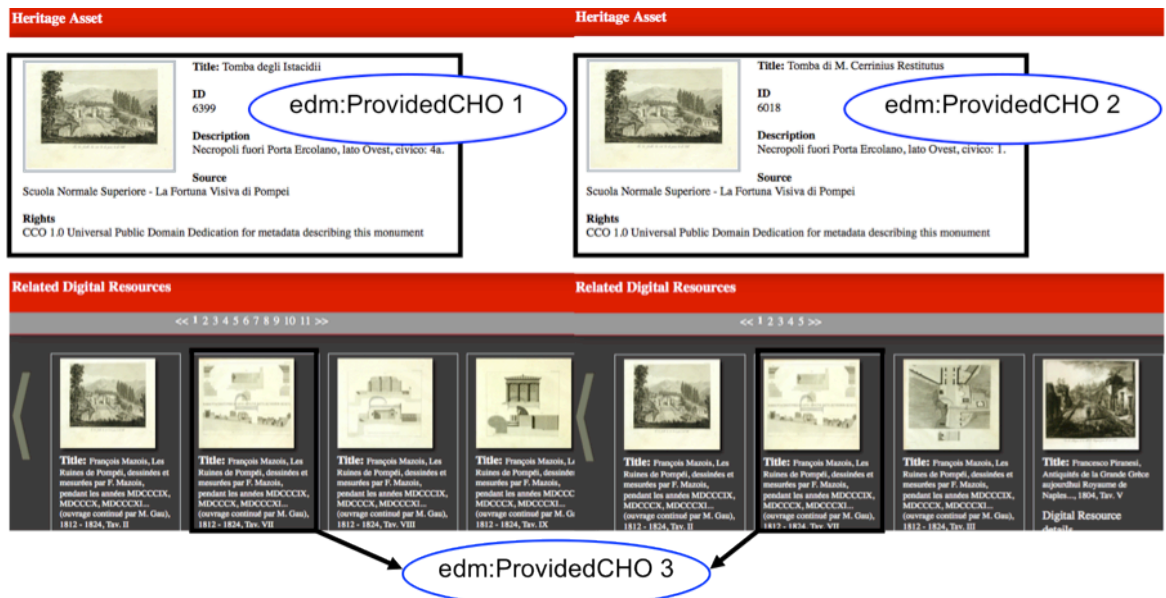

FIG. 5. Scenario 2: some CARARE *Digital Resources* are merged into one edm:ProvidedCHO.
URI ProvidedCHO 1: http://store.carare.eu/landing-page-ha.php?id=iid:2920266&eid=HA:6399
URI ProvidedCHO 2: http://store.carare.eu/landing-page-ha.php?id=iid:2920211&eid=HA:6018

- Scenario 3: a *Heritage Asset* is represented by a series of views of lesser intrinsic cultural relevance (typically, the result of administrative, conservation or cultural resource management activities) which are published online with simple descriptive metadata to accompany a descriptive report. A unique ProvidedCHO is created for the *Heritage Asset* from the descriptive report. In addition edm:WebResources are created for each of the photographs from the identifiers of the Digital *Resources*.

FIG. 6. Scenario 3: mapping CARARE *Digital Resources* as edm:WebResources.
URI ProvidedCHO 1: http://store.carare.eu/uid/iid:3492158/HA:http:/www.kulturarv.dk/fbb/building/7270018

### 4.4 Providing richer and more accurate representations of Web Resources

For each cultural heritage object (a monument, a building or other physical man-made object), CARARE provides multiple digitized resources via the Web. EDM permits detailed description of these resources. One of the requirements of EDM is the separation of information related to a cultural heritage object from the information describing the digital representation of that object. This is particularly important when dealing with rights metadata: an object and its digital representations might have different, or even contradictory, rights statements which determine the conditions for re-use of the content. In addition to the rights information provided in the ore:Aggregation class (see example in Section 4.1), CARARE provides rights information specific to these digital resources:

```
<edm:WebResource
rdf:about="http://iispp.npu.cz/mis_public/documentPreview.htm?id=127767">
  <dc:rights>Národní památkový ústav</dc:rights>
  <dc:rights>2011-01-01</dc:rights>
</edm:WebResource>
```

Note that the rights values described at the WebResource level are also different from the rights that apply to the metadata (metadata delivered by Europeana providers, included CARARE, are available under the Creative Commons CC0 public domain dedication).

### 4.5 Representing and linking to Place entities

The spatial dimension is an important aspect of the archaeological and architectural information. ESE, Europeana's previous metadata format, did not differentiate between temporal and spatial coverage information well enough for archaeological or architectural heritage. EDM, being semantically richer than ESE, allows the representation and the description of place entities by a specific class: edm:Place. CARARE uses it to describe information for a specific place (such as its geo-coordinates) separately from the ProvidedCHO. The link between the ProvidedCHO and related places is represented with Dublin Core's dcterms:spatial.

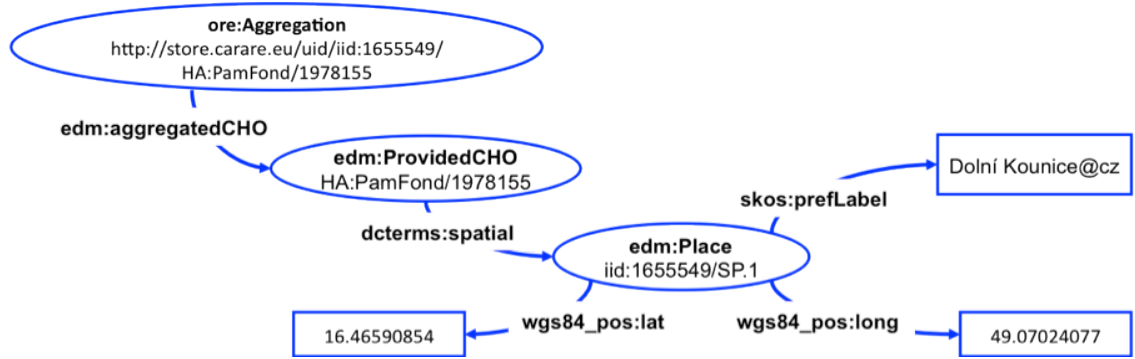

FIG. 7. Representation of a place (edm:Place) related to a Cultural Heritage Object.

```
<edm:Place rdf:about="iid:1655549/SP.1">
  <wgs84_pos:lat>16.46590854</wgs84_pos:lat>
  <wgs84_pos:long>49.07024077</wgs84_pos:long>
  <skos:prefLabel xml:lang="cz">Dolní Kounice</skos:prefLabel>
  <skos:note>132/19, Masarykovo náměstí, Dolní Kounice, 66464, Czech Republic</skos:note>
</edm:Place>
```

Each place has spatial coordinates and a label, sometimes provided in different languages. Spatial coordinates included in CARARE data enable archaeological and architectural datasets to be included in geo-portals alongside other datasets used in planning, development control, tourism and other map-based services. In Europeana, some CARARE datasets have been further enriched with connections to the geospatial dataset Geonames.org available as linked data. All this spatial information can be used to provide new features for end-users, such as Web-based mapping and map browsing of archaeological/architectural sites (CARARE, 2013).

## 5   Related Work

Issues related to mapping (or "crosswalking") between different schemas are not new. One can point to Chan (2006) and Haslhofer (2010) for overviews on schema mapping issues, next to other interoperability problems. With the growing amount and diversity of aggregation services, the challenge has kept receiving continuous attention. It has perhaps even become even more crucial, especially when aggregators involving hundreds of institutes and millions of objects need to align domain-specific schemas to more common ones, at the potential cost of losing data. It becomes then more difficult to overcome general barriers such as the ones identified by (Khoo, 2010). In parallel, the rise of new technological paradigms such as linked data opens new perspectives on the issue (Dunsire, 2011).

We therefore argue that keeping accumulating best practices that can be shared among projects is a meaningful objective. Not only will this help future mapping effort, it can also help the creation and evolution of schemas themselves. Our mapping effort prompted an update to the CARARE schema. Conversely, the creation of EDM by Europeana was prompted by the difficulties Europeana providers were facing when trying to shoehorn their original metadata into the simple ESE format, as seen by comparing reports on representing hierarchical objects in ESE (APEnet, 2012) with the new perspectives EDM opens (HOTF, 2013). Finally, good knowledge of mapping patterns also benefits the design of mapping tools, either for the interface of manual tools (MINT, 2013) or the creation of semi-automatic ones (Walkowska 2012).

In the EDM context, another interesting mapping case comes from the Polymath project (2012). It can be compared with other efforts of mapping library metadata to other schemas, such as the recent BibFrame (2013). In the museum and culture heritage domain, CIDOC CRM, which has significantly inspired EDM as well as the CARARE schema, also comes with extensive literature on crosswalking, such as (Binding, et al., 2006) and (Kakali et al., 2007).

## 6   Discussion and conclusion

The CARARE project worked with the Europeana ingestion team to overcome significant conceptual challenges and to develop a mapping between the CARARE schema and the Europeana Data Model suitable for ensuring homogeneity and quality of archaeological and architectural metadata delivery to Europeana. On this basis, more than two million records of digital resources, representing more than a million monuments and sites, were aggregated in the MoRe system and ingested into Europeana, using the semantically rich EDM, and are now available for public use. By virtue of the particular mapping solutions adopted, users can now discover and browse through a vast spectrum of archaeological and architectural resources from all regions of Europe. They can access immoveable heritage assets, such as archaeological positions, monuments and sites, on the basis of harmonized geographical coordinates where these are available, or historical and current geographic names, and they can visualize geographic relationships between such heritage assets by means of a mapping interface (CARARE, 2013). Using the Europeana results page as a starting point, they can also peruse information on heritage assets in context, or presented on a map, through access to the CARARE landing page that combines metadata on a heritage asset with that of multiple digital resources representing it. Decisions illustrated above were instrumental for the successful integration of information from monument-centric information systems, such as archaeological sites databases, into the web resource-centric world of Europeana.

The mapping effort presented in this paper brought some important lessons for both Europeana and CARARE. Especially, it prompted refinements to the CARARE schema, a process undertaken in the 3D ICONS project alongside enhancement to include paradata and provenance data (d'Andrea & Fernie, 2012). The modeling solutions presented in section 3.3 have been mainly dataset-driven, which could, in the long term, lead to the production of heterogeneous data and the multiplication of mapping profiles. In particular, in Version 2.0 of the CARARE schema includes the following changes:

- The scope of the *Heritage Asset* was broadened to include printed material, archives and born-digital objects relating to the archaeological and architectural heritage;
- *Digital Resource* was simplified to focus on the digital attributes of the resource such as the type, format and location of the digital object.

These changes allow the CARARE schema to be aligned better with EDM. The definitions of *Heritage Asset* and the *Digital Resource* are now closer to the definitions of the edm:ProvidedCHO and the edm:WebResource, which in the future will make the mapping task easier and will enhance the interoperability between CARARE and Europeana.

In the future, we plan to extend EDM with elements derived from the CARARE XML schema (considering the CARARE schema as an application profile of EDM), so that it becomes more suitable for expressing RDF data for archaeological and architectural information aggregation.